\documentclass[a4paper,11pt]{article}
\pdfoutput=1 
\usepackage{jheppub}

\usepackage{multirow, graphicx,amssymb,url,mathrsfs,amsmath}
\usepackage{wrapfig,boxedminipage,subfigure,epsfig}
\usepackage{amsxtra,amstext,latexsym,dsfont,amsfonts}
\usepackage{color}
\usepackage[dvipsnames]{xcolor}
\usepackage{float}
\usepackage{slashed}
\usepackage{calligra}
\DeclareFontShape{T1}{calligra}{m}{n}{<->s*[2.2]callig15}{}
\DeclareMathAlphabet{\mathcalligra}{T1}{calligra}{m}{n}

\definecolor{shadecolor}{rgb}{0.9,0.9,0.95}

\newcommand{\be}{\begin{equation}}
\newcommand{\ee}{\end{equation}}
\newcommand{\bea}{\begin{eqnarray}}
\newcommand{\eea}{\end{eqnarray}}

\newcommand{\dd}{\mathrm{d}}

\title{\Huge \centering On pole-skipping with gauge-invariant variables in holographic axion theories}

\author[a,b,c]{Yongjun Ahn,}
\author[d]{Viktor Jahnke,}
\author[e,f]{Hyun-Sik Jeong,}
\author[d]{Chang-Woo Ji,}
\author[d,g]{Keun-Young Kim}
\author[h]{and Mitsuhiro Nishida}

\preprint{IFT-UAM/CSIC-24-23}

\affiliation[a]{School of Physics and Astronomy, Shanghai Jiao Tong University, Shanghai 200240, China}
\affiliation[b]{Wilczek Quantum Center, School of Physics and Astronomy, Shanghai Jiao Tong University, Shang-
hai 200240, China}
\affiliation[c]{Shanghai Research Center for Quantum Sciences, Shanghai 201315, China}
\affiliation[d]{Department of Physics and Photon Science, Gwangju Institute of Science and Technology,\\123 Cheomdan-gwagiro, Gwangju 61005, Korea}
\affiliation[e]{Instituto de F\'isica Te\'orica UAM/CSIC, Calle Nicol\'as Cabrera 13-15, 28049 Madrid, Spain}
\affiliation[f]{Departamento de F\'isica Te\'orica, Universidad Aut{\'o}noma de Madrid, 28049 Madrid, Spain}
\affiliation[g]{Research Center for Photon Science Technology, Gwangju Institute of Science and Technology, 123 Cheomdan-gwagiro, Gwangju 61005, Korea}
\affiliation[h]{Department of Physics, Pohang University of Science and Technology, Pohang 37673, Korea}

\emailAdd{yongjunahn@sjtu.edu.cn}
\emailAdd{viktorjahnke@gist.ac.kr}
\emailAdd{hyunsik.jeong@csic.es}
\emailAdd{physianoji@gm.gist.ac.kr}
\emailAdd{fortoe@gist.ac.kr}
\emailAdd{nishida124@postech.ac.kr}

\abstract{
We study the pole-skipping phenomenon within holographic axion theories, a common framework for studying strongly coupled systems with chemical potential ($\mu$) and momentum relaxation ($\beta$). Considering the backreaction characterized by $\mu$ and $\beta$, we encounter coupled equations of motion for the metric, gauge, and axion field, which are classified into spin-0, spin-1, and spin-2 channels. Employing gauge-invariant variables, we systematically address these equations and explore pole-skipping points within each sector using the near-horizon method. Our analysis reveals two classes of pole-skipping points: {\it regular and singular pole-skipping points} in which the latter is identified when standard linear differential equations exhibit singularity. Notably, pole-skipping points in the lower-half plane are regular, while those elsewhere are singular. This suggests that the pole-skipping point in the spin-0 channel, associated with quantum chaos, corresponds to a singular pole-skipping point. Additionally, we observe that the pole-skipping momentum, if purely real or imaginary for $\mu=\beta=0$, retains this characteristic for $\mu \neq0$ and $\beta \neq 0$.
}

\begin{document}
\maketitle
\section{Introduction}
The holographic correspondence~\cite{Maldacena:1997re,Witten:1998qj,Witten:1998zw,Gubser:1998bc}, also known as holography or AdS/CFT correspondence, has become instrumental in the study of strongly interacting many-body quantum systems through the lens of retarded Green's functions~\cite{Ammon:2015wua,Hartnoll:2016apf,Hartnoll:2009sz,Zaanen:2015oix,Baggioli:2019rrs}, which characterize the system's near-equilibrium behavior.
In particular, in recent years, studies in the framework of AdS/CFT correspondence have revealed a remarkable and universal property associated with Green's functions {of holographic systems}, known as pole-skipping~\cite{Grozdanov:2017ajz,Blake:2017ris,Blake:2018leo}.

Pole-skipping is a phenomenon that takes place when the dual boundary Green's function in the complex momentum space ($\omega, k$) becomes ill-defined or multi-valued at a special point ($\omega_\star, k_\star$), called pole-skipping point. Within the context of holography, the pole-skipping phenomenon can be elucidated by studying the equations of motion of bulk fields in the vicinity of black hole horizons, as the dual Green's functions are linked to bulk fluctuations.

Following standard holographic techniques (e.g.~\cite{Kovtun:2005ev}), Green's functions are determined by solving the bulk equations of motion in complex Fourier space and imposing specific boundary conditions at the AdS boundary and black hole horizon. Since the equations of motion are typically second-order differential equations, two linearly independent solutions arise, which can be taken as satisfying ingoing and outgoing boundary conditions at the horizon. Generally, imposing ingoing boundary conditions uniquely specifies (up to an overall factor) the solution and corresponding boundary Green's function.  However, at the pole-skipping point $(\omega_\star,k_\star)$, imposing ingoing boundary conditions is no longer sufficient to guarantee the uniqueness of the solution because both solutions become ingoing at the horizon, rendering Green's function non-uniquely defined~\cite{Grozdanov:2017ajz, Blake:2018leo, Grozdanov:2019uhi, Blake:2019otz, Natsuume:2019xcy}. The non-uniqueness of the ingoing bulk solution at the horizon offers a straightforward method for identifying pole-skipping points ~\cite{Blake:2018leo, Blake:2019otz}, termed the {\it near-horizon method}.

The phenomenon of pole-skipping was initially discovered in the study of the many-body quantum chaos within the framework of holography~\cite{Grozdanov:2017ajz, Blake:2017ris, Blake:2018leo}. 
Within this context, the pole-skipping point of the energy density Green's function is associated with both the Lyapunov exponent ($\lambda_L$) and butterfly velocity ($v_B$), which characterize the behavior of out-of-time-ordered correlators (OTOCs).\footnote{Pole-skipping has also been theoretically established as a universal prediction within the framework of an effective field theory for maximally chaotic systems~\cite{Blake:2017ris, Blake:2021wqj}.} Therefore, the pole-skipping phenomenon can in principle provide an approach to directly extract quantum chaos properties ($\lambda_L, v_B$) from the energy density Green's two-point function, without the need for computing OTOCs. 
Because of its connection to quantum chaos, the pole-skipping phenomenon has been extensively investigated from both physical and mathematical perspectives in various scenarios~\cite{Blake:2018leo, Grozdanov:2019uhi, Blake:2019otz, Natsuume:2019xcy, Natsuume:2019sfp, Natsuume:2019vcv, Ceplak:2019ymw, Ahn:2019rnq, Ahn:2020bks, Gu:2016oyy, Haehl:2018izb, Abbasi:2020xli, Abbasi:2020ykq, Liu:2020yaf, Ramirez:2020qer, Ahn:2020baf, Natsuume:2020snz, Ceplak:2021efc, Jeong:2021zhz, Natsuume:2021fhn, Blake:2021hjj, Jeong:2022luo, Wang:2022mcq, Amano:2022mlu, Yuan:2023tft, Grozdanov:2023txs, Natsuume:2023lzy, Ning:2023pmz, Grozdanov:2023tag, Jeong:2023rck, Natsuume:2023hsz, Abbasi:2023myj, Jeong:2023ynk, Yadav:2023hyg, Baishya:2023ojl, Atashi:2022emh}. 
The exemplary of its implications include, for example, a universal bound on transport coefficients (such as diffusivity) in many-body systems~\cite{Blake:2018leo, Jeong:2021zhz, Jeong:2022luo, Jeong:2023ynk}, and the reconstruction and constraints of collective excitations (i.e., quasi-normal modes)~\cite{Abbasi:2020xli, Grozdanov:2023tag, Jeong:2023ynk}. Early studies of pole-skipping phenomena from the point of view of the boundary theory include pioneering works in SYK chains~\cite{Gu:2016oyy} and two-dimensional CFTs with large central charges~\cite{Haehl:2018izb}.

In this work, we systematically investigate the pole-skipping phenomenon in a five-dimensional Einstein-Maxwell theory coupled with massless scalar fields (`axions').  Specifically, we focus on analyzing the linear axion model proposed in~\cite{Andrade:2013gsa}, which offers a holographic\footnote{For recent developments regarding pole-skipping phenomena in non-holographic systems or non-black hole backgrounds, see \cite{Natsuume:2021fhn, Natsuume:2023lzy, Natsuume:2023hsz}.} framework for studying strongly coupled systems at finite chemical potential and in the presence of momentum relaxation.
This model is particularly appealing due to its closed-form analytical background featuring broken translational symmetry.\footnote{The linear-axion model's symmetry-breaking pattern can be manipulated by adjusting the asymptotic boundary condition of the scalar field~\cite{Baggioli:2021xuv}. For a similar analysis involving vector fields to explore the dynamic electromagnetism in the dual boundary field theory, refer to \cite{Ahn:2022azl,Jeong:2023las}.} It has yielded noteworthy and celebrated outcomes in applied holography, especially in the investigation of anomalous transport properties and collective dynamics of strongly coupled phases~\cite{Davison:2013txa,Davison:2014lua,Gouteraux:2014hca,Blake:2016jnn,Blake:2017qgd,Baggioli:2017ojd,Ahn:2017kvc,Blauvelt:2017koq,Alberte:2017cch,Jeong:2018tua,Davison:2018ofp,Blake:2018leo,PhysRevLett.120.171602,Ammon:2019wci,Ahn:2019lrh,Jeong:2019zab,Arean:2020eus,Jeong:2021wiu,Liu:2021qmt,Jeong:2021zhz,Wu:2021mkk,Jeong:2021zsv,Huh:2021ppg,Baggioli:2022pyb,Baggioli:2022uqb,Ahn:2023ciq,Jeong:2023ynk}. Quantum information applications can also be explored in \cite{RezaMohammadiMozaffar:2016lbo,Yekta:2020wup,Li:2019rpp,Zhou:2019xzc,Huang:2019zph,Jeong:2022zea,HosseiniMansoori:2022hok}. Furthermore, the model recently finds application in the AdS/Deep learning correspondence in \cite{Ahn:2024gjf}. For an in-depth and up-to-date review of the linear-axion model and an extensive list of references, we refer to \cite{Baggioli:2021xuv}.\\

Below, we provide additional motivations for investigating pole-skipping in linear axion models, along with an explanation of our choice of model.

\paragraph{Backreaction effects of the gauge and axion fields.}
In the linear axion model, ~\cite{Andrade:2013gsa}, the parameters $\beta$ and $\mu$ control the backreaction of the axion and gauge field on the geometry, respectively. $\beta$ serves as the momentum relaxation parameter, while $\mu$ acts as the chemical potential in the dual boundary theory.  When  $\mu \neq 0$ or $\beta \neq 0$, only a limited number of pole-skipping points, such as the leading pole-skipping point in the spin-0 channel, have been thoroughly investigated in previous literature, e.g., \cite{Blake:2018leo, Jeong:2021zhz, Jeong:2022luo, Abbasi:2020ykq}. Our work represents the first comprehensive examination considering both finite $\mu$ and $\beta$, offering a more complete analysis of the pole-skipping phenomenon in linear axion models.

\paragraph{Regarding the prospective physical implications.}
Furthermore, the effects of $\beta$ and $\mu$ are significant in physical applications related to pole-skipping. Particularly, $\beta$ plays a crucial role in the context of the gravitational sound mode within the spin-0 channel, as highlighted in previous works~\cite{Blake:2018leo, Jeong:2021zhz, Jeong:2022luo, Jeong:2023ynk}. In this context, the celebrated universal {lower} bound of {the diffusion constant} in holography~\cite{Blake:2016wvh, Blake:2016sud}  can be understood through the pole-skipping phenomena, particularly in the regime of strong momentum relaxation.\footnote{This universal bound has been extensively explored and discussed in literature, see, e.g.,~\cite{Lucas:2018wsc, Davison:2018ofp, Gu:2017njx, Ling:2017jik, Gu:2017ohj, Blake:2016jnn, Blake:2017qgd, Wu:2017mdl, Li:2019bgc, Ge:2017fix, Li:2017nxh, Ahn:2017kvc, Baggioli:2017ojd, Kim:2017dgz, Aleiner:2016aa, Patel:2017vfp, Patel:2016wdy, Bohrdt:2016vhv, Werman:2017abn, Chen:2019dch}.} Therefore, it becomes imperative to examine the role of momentum relaxation (along with finite charge) across all possible channels, spanning from spin-0 to spin-2 sectors.\footnote{It is also worth noting that pure Schwarzschild black geometries without backreaction of matter fields may not always be suitable for understanding many-body systems in holography.  A notable example can be found in the study of holographic superfluids~\cite{Hartnoll:2008vx, Hartnoll:2008kx}, where it becomes evident that the probe limit analysis {sometimes} lacks consistency with the scenario involving backreaction, especially far from the critical temperature.
}

\paragraph{Near-horizon analysis in terms of gauge-invariant variables.} 
The finite backreaction of the gauge and axion field on the geometry introduces technical difficulties in the analysis of bulk fluctuations, which are absent when $\beta=\mu=0$. For instance, a finite charge couples the gauge field fluctuations with the metric field fluctuations, demanding a more sophisticated approach. For such a case, the use of gauge-invariant variables becomes useful (and sometimes indispensable), particularly when investigating collective excitations (i.e., poles of Green's functions). In this work, we introduce a comprehensive methodology for the determination of pole-skipping points using gauge-invariant variables.\\

 It is worth noting that a prior study~\cite{Abbasi:2020ykq} investigated pole-skipping in a similar model. However, they only considered the chemical potential effect without momentum relaxation. Additionally, their examination focused on identifying the leading pole-skipping point in the spin-0 channel and did not involve using gauge-invariant variables. Consequently, our work can also be regarded as an exhaustive extension of~\cite{Abbasi:2020ykq}, encompassing all possible channels and accounting for the presence of momentum relaxation.\\

This paper is organized as follows.
In Section \ref{sec:EOM}, we provide the equations of motion governing bulk fluctuations across all channels in the presence of backreaction from the gauge and axion field. These equations are derived in terms of gauge-invariant variables. In Section \ref{sec:pspwithbackract}, we apply a near-horizon analysis to the aforementioned equations of motion to determine the pole-skipping points. Section \ref{sec:discussion} is devoted to conclusions. 
In Appendix \ref{sec:zero}, for completeness, we discuss additional singular cases which in general do not lead to the appearance of pole-skipping points. In Appendix \ref{sec:EOMcoef}, we give the explicit form of some auxiliary functions that were used to write the equations of motion in Section \ref{sec:EOM}.\\

\section{Background geometry and equations of motion } \label{sec:EOM}

\subsection{Background equations of motion} \label{subsec:SetupwithouBR}
We consider the linear axion model proposed in~\cite{Andrade:2013gsa} in $(4+1)$ dimensions\footnote{To perform a systematic analysis of the pole-skipping phenomenon, including fluctuations in the spin-0, spin-1, and spin-2 channels, we consider a model in (4+1) bulk dimensions. Note that the spin-2 channel of metric fluctuations is absent in lower dimensional models.}, namely
\begin{equation}
\label{eq:action}
S=\int \dd^{5}x \sqrt{-g} \left( R+\frac{12}{L^2}-\frac{1}{2}\sum_I^{3}(\partial \chi_I)^2-\frac{1}{4}F^2 \right),
\end{equation}
where $\chi_I$ denotes massless scalar fields (axions), and $F=\dd A$ is the electromagnetic tensor. 

The action (\ref{eq:action}) admits asymptotically AdS solutions in which $L$ defines the AdS length scale. From now on, we set $L=1$ without loss of generality. Working in coordinates in which the AdS boundary is located at $r\rightarrow \infty$, the authors of~\cite{Andrade:2013gsa} show that the model (\ref{eq:action}) admits solutions in which the metric takes the form
\begin{equation} \label{eq:BackMetric}
\dd s^{2} = -r^{2} f(r) \dd t^{2} + \frac{\dd r^{2}}{r^{2} f(r)} + r^2 \dd \vec{x}^{2}\,,
\end{equation}
while the axion fields $\chi_I$ and gauge field $A$ are given by
\begin{equation} \label{eq:BackGaugeAxion}
\chi_I = \beta \delta_{Ia}x^{a}\,, \qquad A = A_t(r) \dd t\,.
\end{equation}

\paragraph{Background solution.}
Plugging (\ref{eq:BackMetric}) and (\ref{eq:BackGaugeAxion}) in the equations of motion resulting from (\ref{eq:action}), one finds
\begin{equation}\label{eq:givenBackground}
\begin{split}
    f(r)=\left( 1 - \frac{r_h^2}{r^2} \right) \left( 1 + \frac{r_h^2}{r^2}-\frac{\beta^2}{4 r^2}+\frac{\mu^2 r_h^2}{3 r^4} \right)\,, \qquad
    A_t(r)=\mu\left(1-\frac{r_h^{2}}{r^{2}}\right)\,.
\end{split}
\end{equation}
where $r_h$ is the horizon radius, while $\beta$ and $\mu$ are interpreted as momentum relaxation parameter and chemical potential of the system in the dual boundary description. The Hawking temperature is given by
\begin{equation}\label{eq:Temp}
    T=\frac{r_h^2 f'(r_h)}{4 \pi} = \frac{r_h}{\pi}-\frac{\beta^2}{8 \pi r_h}-\frac{\mu^2}{6\pi r_h}\,.
\end{equation}
For later convenience, we introduce the re-scaled horizon radius $\bar{r}_h$ as follows
\begin{equation}\label{eq:horizonRadius}
    \bar{r}_h:=\frac{r_h}{2\pi T}=\frac{1}{4} + \sqrt{\frac{1}{16} + \frac{\bar{\beta}^2}{8} + \frac{\bar{\mu}^2}{6}}\,,
\end{equation}
as well as the re-scaled momentum relaxation parameter $\bar{\beta}=\beta / (2\pi T)$ and chemical potential $\bar{\mu}=\mu / (2\pi T)$.  

\paragraph{Eddington-Finkelstein coordinates.}
To compute the retarded Green's function $G^R_{\mathcal{O}\mathcal{O}}$ holographically, one needs to consider fluctuations of the correspondent bulk field $\psi$ satisfying ingoing boundary conditions at the horizon and study the near boundary behavior of $\psi$. To find the ingoing fluctuations, it is convenient to use the ingoing Eddington-Finkelstein coordinates
\begin{equation}
    v=t+r_*, \qquad \dd r_*=\frac{1}{r^2 f(r)}\dd r\,.
\end{equation}
in terms of which the metric \eqref{eq:BackMetric} becomes
\begin{equation}
\label{co:inEF}
\dd s^2=-r^2f(r) \dd v^2 + 2\dd v\,\dd r + r^2 \dd \vec{x}^2\,.
\end{equation}
In these coordinates, imposing ingoing boundary condition at the horizon guarantees that the solutions are regular.

\subsection{Equations of motion for the fluctuations} \label{sec:fluc}
We consider fluctuations of the metric, gauge, and axion field and write them in terms of plane waves propagating along the $z$-direction
\begin{align}
\begin{split}
    \delta g_{\mu\nu}  = \delta g_{\mu\nu}(r) e^{-i(\omega v - k z)}\,,\qquad
    \delta A_\mu  = \delta A_\mu(r) e^{-i(\omega v - k z)}\,, \qquad
    \delta \chi_I  = \delta \chi_I(r) e^{-i(\omega v - k z)}\,.
\end{split}
\end{align}
After selecting fluctuations propagating in the $z$-direction, the remaining world-volume symmetry group is $O(2)$. We can then use this group to categorize the fluctuations based on their symmetry properties. In the radial gauge, we find:
\begin{equation}\label{templabelsetup}
\begin{aligned}
    & \text{Spin-0 channel:}   &\quad &\left({\delta g}_{vv},\, {\delta g}_{vz},\, {\delta g}_{zz},\, \mathfrak{g},\, {\delta A}_{v},\, {\delta A}_{z},\, {\delta \chi}_{z}\right)\,, \\
    & \text{Spin-1 channel:}   &\quad &\left({\delta g}_{v\alpha},\, {\delta g}_{z\alpha},\, {\delta A}_{\alpha},\, {\delta \chi}_{\alpha}\right)\,, \\
    & \text{Spin-2 channel:}   &\quad &\,\,\, \delta g_{\alpha \beta} - \delta_{\alpha \beta} \frac{\mathfrak{g}}{2}\,,
\end{aligned}
\end{equation}
where $\mathfrak{g} := \sum_\alpha \delta g_{\alpha \alpha}$, and $\alpha=x,\,y$. For each spin channel, we find gauge-invariant variables and the corresponding coupled equations of motion following the method described in \cite{Abbasi:2020ykq}.

\subsubsection{Spin-$0$ channel}
For the spin-$0$ channel, the gauge-invariant variables are found as 
\begin{equation}\label{eq:GaugeInvSpin0}
\begin{aligned}
    & Z_z(r)=\frac{1}{r^2}\left(2\frac{\omega}{k} {\delta g}_{vz} + {\delta g}_{vv} + \frac{\omega^2}{k^2} {\delta g}_{zz} - \left( \frac{\omega^2}{k^2} - f(r) - \frac{1}{2} r f'(r) \right) \frac{\mathfrak{g}}{2} \right)\,,
    \\
    & E_z(r) = \frac{{\delta A}_{v}}{r_h^2} + \frac{\omega}{k} \frac{{\delta A}_{z}}{r_h^2} - \frac{\mu}{r^4}\frac{\mathfrak{g}}{2}\,,
    \\
    & \Phi_z(r) = 2k {\delta \chi}_{z} + \frac{i \beta}{r^2}\left({\delta g}_{zz} - \frac{\mathfrak{g}}{2} \right)\,.
\end{aligned}
\end{equation}
In terms of these variables, the equations of motion can be written in terms of three coupled differential equations:

\begin{align} \label{eq:Spin0EOM}
\begin{split}
    & Z_z''+\left( \frac{5}{r} + \frac{r^2 f'-2 i \omega}{r^2 f} -\frac{2 H_2'}{H_2} + \beta^2 \frac{B_1}{H_4} - \mu^2\frac{4 k^2 r_h^4}{3 r^7 H_1} + \beta^2 \mu^2 \frac{B_2}{H_3 H_5} \right)Z_z' \\
    & \qquad \qquad -\frac{1}{f}\left(\frac{k^2+\beta^2+3 i r \omega}{r^4} - \frac{2 i \omega H_2'}{r^2 H_2} + \frac{k^4 f'^2}{3 H_2}  - \frac{\mu^2 B_3 + \beta^2 B_4}{H_2 H_3} \right)Z_z  \\
    & \qquad \qquad \qquad \qquad + \frac{\mu}{H_3}\left( B_5 E_z' - \frac{B_6}{r^2 f}E_z \right) + \frac{\beta}{k^2 H_3}\left( B_7 \Phi_z' - \frac{B_8}{r^2 f} \Phi_z \right) = 0\,, \\
    & E_z'' + \left(\frac{3}{r} + \frac{r^2 f' - 2 i \omega}{r^2 f} - \frac{H_1'}{H_1} + \mu^2 \frac{4 k^2 r_h^4}{3 r^7 H_1 } + \mu^2 \beta^2 \frac{k^2 f C_1}{H_1 H_3} \right) E_z'  \\
    & \qquad \qquad - \left( \frac{1}{f} \left( \frac{k^2 + i r \omega}{r^4} - \frac{i \omega H_1' }{r^2 H_1 } - \frac{\mu^2 C_2}{H_1 H_6} \right) - \frac{\mu^2 \beta^2 C_3}{H_1 H_3 H_6} \right) E_z  \\
    & \qquad \qquad \qquad \qquad + \frac{\mu}{H_3}\left( C_4  Z_z'  + \frac{C_5}{r^2 f} Z_z \right) + \frac{\mu \beta}{k^2 H_3} \left( C_6 \Phi_z' + \frac{C_7}{r^2 f} \Phi_z \right)
    = 0\,, \\
    & \Phi_z'' +\left( \frac{5}{r} + \frac{r^2 f'  - 2 i \omega}{r^2 f } \right) \Phi_z'  - \frac{k^2 + \beta^2 + 3 i r \omega}{r^4 f }\Phi_z  = 0\,,
\end{split}
\end{align}
where the primes denote derivatives with respect to $r$. Auxiliary functions $H_n, B_n$, and $C_n$ ($n=1,2,3,\cdots$) are functions of $r$. Among them, $H_n$ are particularly important because they determine the singularity structure of pole-skipping points, so we show their expressions in Appendix \ref{sec:EOMcoef}. However, we do not display the explicit expressions of $B_n$ and $C_n$ because they are too complicated and not very illuminating.

\subsubsection{Spin-$1$ channel} \label{BREOMVectorMode}
For the spin-$1$ channel, the gauge-invariant variables are found as 
\begin{equation} \label{eq:GaugeInvSpin1}
\begin{aligned}
    & Z_\alpha(r)=\frac{1}{r^2} \left( {\delta g}_{v \alpha} + \frac{\omega}{k} {\delta g}_{z \alpha} \right)\,,
    \\
    & E_\alpha(r) = \frac{{\delta A}_{\alpha}}{r_h^2}\,,
    \\
    & \Phi_\alpha(r) = k {\delta \chi}_{\alpha} + i \beta \frac{{\delta g}_{z\alpha}}{r^2}\,,
\end{aligned}
\end{equation}
where $\alpha =x,\,y$. In terms of these variables, the equations of motion can be written in terms of three coupled differential equations:
\begin{align}\label{eq:Spin1EOM}
\begin{split}
    & {Z_{\alpha}}''  +   \left( \frac{5}{r}+\frac{r^2 f'-2 i \omega}{r^2 f} - \frac{H_7'}{H_7} \right){Z_{\alpha}}'
    -\frac{1}{f} \left( \frac{k^2 + \beta^2 + 3 i r \omega}{r^4} -\frac{i \omega H_7'}{r^2 H_7} \right) Z_{\alpha}
    \\
    & \hspace{5.5cm}
    +\mu \frac{2 {r_h}^4}{r^5} \left( {E_\alpha}' + \left( \frac{-i \omega}{r^2 f}+\frac{\omega^2}{{H_7}}\frac{f'}{f}\right) {E_\alpha} \right)
    \\
    & \hspace{7.9cm}
    +\beta\frac{\omega}{k}\frac{i f'}{H_7} \left( {\Phi_\alpha}'-\frac{i \omega}{r^2 f}\Phi_\alpha \right)
    =0\,,
    \\
    & E_\alpha''
    + \left(\frac{3}{r} + \frac{r^2 f'-2 i \omega}{r^2 f} \right) E_\alpha'
    -\frac{1}{f} \left( \frac{k^2+i r \omega}{r^4} + \frac{4 \mu^2 r_h^4 \omega^2}{r^8 H_7} \right) E_\alpha
    \\
    & \hspace{5.5cm}
    -\mu\frac{2 (k^2+\beta^2)}{r^3 H_7}\left( Z_{\alpha}'- \frac{i \omega}{r^2 f} Z_{\alpha} \right)
    \\
    & \hspace{7.5cm}
    -\mu\beta\frac{\omega}{k}\frac{2 i}{r^3 H_7}\left( \Phi_\alpha'-\frac{i \omega}{r^2 f}\Phi_\alpha \right)
    = 0\,,
    \\
    & \Phi_\alpha'' +\left(\frac{5}{r}+\frac{r^2 f' - 2 i \omega}{r^2 f} \right)\Phi_\alpha' - \frac{k^2 + \beta^2 + 3 i r \omega }{r^4 f}\Phi_\alpha
    =0\,,
\end{split}
\end{align}
where the function $H_7$ is written in Appendix \ref{sec:EOMcoef}.

\subsubsection{Spin-$2$ channel}
For the spin-$2$ channel, the gauge-invariant variables are written based only on metric fluctuations as 
\begin{equation}
\begin{aligned}
    Z_{\alpha\beta}(r)=\frac{1}{r^2}\left(\delta g_{\alpha \beta} - \delta_{\alpha \beta} \frac{\mathfrak{g}}{2} \right)\,,
\end{aligned}
\end{equation}
where $\alpha,\, \beta$ denote transverse spatial coordinates such as $x,\,y$. In terms of these variables, one finds the following decoupled equation of motion:
\begin{equation}\label{eq:Spin2EOM}
\begin{aligned}
    & Z_{\alpha\beta}'' + \left( \frac{5}{r} + \frac{r^2 f' - 2 i \omega}{r^2 f} \right)Z_{\alpha\beta}'
    \\
    & \hspace{1.5cm}
    - \frac{1}{f} \left( \frac{k^2 + \beta^2 + 3 i r \omega}{r^4} + \frac{12(f-1)}{r^2} + \frac{8 f'}{r} + f'' + \beta^2 \frac{1}{2 r^4} - \mu^2 \frac{2 r_h^4 }{r^8} \right) Z_{\alpha\beta}
    =0\,.
\end{aligned}
\end{equation}

\section{Pole-skipping in linear axion models} \label{sec:pspwithbackract}

The pole-skipping points can be found by solving the equations of motion in the near-horizon region following the so-called {\it near-horizon method}~\cite{Blake:2018leo, Blake:2019otz}. In this section, we carefully show how to apply the near-horizon method to compute pole-skipping points in the presence of a non-trivial backreaction of axion and gauge fields on the geometry. As shown in Sec.~\ref{sec:EOM}, the main effect of such backreaction is to couple the equations of motion in each sector. We also compute pole-skipping points considering singular cases.

\subsection{Near-horizon method with gauge-invariant variables} \label{section:Frobenius method2}
Pole-skipping points are special values of $\omega$ and $k$ at which there is more than one solution to the equations of motion satisfying ingoing boundary conditions at the horizon. To find ingoing solutions that are regular around the black hole horizon, we employ the Frobenius method and write the gauge-invariant variables $Z$, $E$, $\Phi$ as follows:
\begin{equation} \label{eq:ZEPhiAnsatz}
    Z=\sum_{i=0}^{\infty}Z^{(i)}(r-r_h)^{i}\,, \qquad
    E=\sum_{i=0}^{\infty}E^{(i)}(r-r_h)^{i}\,, \qquad
    \Phi=\sum_{i=0}^{\infty}\Phi^{(i)}(r-r_h)^{i}\,.
\end{equation}
Substituting these expansions into the coupled equations of motion \eqref{eq:Spin0EOM} or \eqref{eq:Spin1EOM}, each equation takes the following near-horizon form:
\begin{equation} \label{eq:NearHorizon Ansatz}
\begin{aligned}
    S_{Z}=\sum_{i=1}^{\infty}S_{Z}^{(i)}(r-r_h)^{i-2}=0\,,\\
    S_{E}=\sum_{i=1}^{\infty}S_{E}^{(i)}(r-r_h)^{i-2}=0\,,\\
    S_{\Phi}=\sum_{i=1}^{\infty}S_{\Phi}^{(i)}(r-r_h)^{i-2}=0\,,
\end{aligned}
\end{equation}
where $S_{\psi}$ denotes a second order differential equation containing a $\psi''$ term, with $\psi = Z, E,$ and $\Phi$. The coefficients $S_\psi^{(i)}$ are computed as
\begin{equation}\label{eq:BREOMSeries} \footnotesize
\begin{aligned}
    &S_{Z}^{(1)}=-\left( M_{11} Z^{(0)} + M_{12}E^{(0)} + M_{13}\Phi^{(0)} + (i\bar{\omega}-1)Z^{(1)} \right)\,,\\
    &S_{E}^{(1)}=-\left( M_{21} Z^{(0)} + M_{22}E^{(0)} + M_{23}\Phi^{(0)} + (i\bar{\omega}-1)E^{(1)} \right)\,,\\
    &S_{\Phi}^{(1)}=-\left(M_{33}\Phi^{(0)} + (i\bar{\omega}-1)\Phi^{(1)} \right)\,,\\
    &\vdots\\
    &S_{Z}^{(n)}=-n \left( \cdots  + M_{(3n-2)(3n-2)}\phi^{(n-1)} + M_{(3n-2)(3n-1)}E^{(n-1)} + M_{(3n-2)3n}\Phi^{(n-1)} + (i\bar{\omega}-n)Z^{(n)} \right)\,,\\
    &S_{E}^{(n)}=-n \left( \cdots + M_{(3n-1)(3n-2)}\phi^{(n-1)} + M_{(3n-1)(3n-1)}E^{(n-1)} + M_{(3n-1)3n}\Phi^{(n-1)} + (i\bar{\omega}-n)E^{(n)} \right)\,,\\
    &S_{\Phi}^{(n)}=-n \left( \cdots + M_{3n3n}\Phi^{(n-1)} + (i\bar{\omega}-n)\Phi^{(n)} \right)\,.\\
\end{aligned}
\end{equation}
The ansatz \eqref{eq:ZEPhiAnsatz} solves the equations of motion if all the coefficients $S_\psi^{(i)}$ appearing in \eqref{eq:BREOMSeries} vanish. This condition can be written as follows:
\begin{equation} \label{eq:MatrixEqn}
\mathcal{M} \Psi \equiv
\begin{pmatrix}
M_{11} & M_{12} & M_{13} & i \bar{\omega}-1 &  &  & & & & \\
M_{21} & M_{22} & M_{23} &  &  i \bar{\omega}-1 &  & & & &\cdots\\
&  & M_{33} &  &  &  i \bar{\omega}-1 & & & & \\
M_{41} & M_{42} & M_{43} & M_{44} & M_{45} & M_{46} & i \bar{\omega}-2 &  & & \\
M_{51} & M_{52} & M_{53} & M_{54} & M_{55} & M_{56} &  &  i \bar{\omega}-2 &  & \cdots\\
&  & M_{63} & & & M_{66} &  &  &  i \bar{\omega}-2 & \\
&  & \vdots & & & \vdots & & & & \ddots
\end{pmatrix}
\begin{pmatrix}
Z^{(0)} \\
E^{(0)} \\
\Phi^{(0)} \\
Z^{(1)} \\
E^{(1)} \\
\Phi^{(1)} \\
\vdots
\end{pmatrix}
=
0\,,
\end{equation}
where $\bar{\omega}=\omega / (2\pi T)$, $\bar{k}=k / (2\pi T)$, and $M_{ij}$ is generally a function of $\bar{\omega}$ and $\bar{k}$. In general, at non-special points $(\bar{\omega},\, \bar{k})$, all the coefficients $\{Z^{(i)},\, E^{(i)},\, \Phi^{(i)} \}$ are determined by the free parameters $\{Z^{(0)},\, E^{(0)},\, \Phi^{(0)} \}$, and the solutions are unique up to an overall factor. However, there are special points $(\bar{\omega}_{\star}, \bar{k}_{\star})$ at which additional free parameters appear, and the solutions are no longer unique. These special points are called {\it pole-skipping points}, and they are distinguished into two types: {\it regular} pole-skipping points and {\it singular} pole-skipping points. The regular(singular) pole-skipping point is defined as a pole-skipping point of which equations of motion are regular(ill-defined) near the horizon. Therefore, some of the matrix components $M_{ij}$ in \eqref{eq:MatrixEqn} can be singular at the singular pole-skipping points while $M_{ij}$ is regular at the regular pole-skipping points.

The way to identify the regular pole-skipping points is the following:
\begin{equation} \label{eq:omegaDetM}
    i \bar{\omega}_{\star}= n, \qquad \left.\det (\mathcal{M}_{n})\right|_{(\bar{\omega}_{\star}, \bar{k}_{\star})}=0\,,
\end{equation}
where $n$ is a natural number and $\mathcal{M}_{n}$ a square matrix which is taken from the matrix $\mathcal{M}$ in \eqref{eq:MatrixEqn} up to the $3n$-th column. Note that the pole-skipping frequencies take the same value as the imaginary Matsubara frequencies $\omega_n = -i 2 \pi T n$~\cite{Blake:2019otz,Ceplak:2021efc}.

The singular pole-skipping point can be analyzed after parameterizing the way to approach the singular point $(\bar{\omega}_\star, \bar{k}_\star)$ by $s$ as follows:
\begin{equation} \label{eq:singp-sAnalDef}
    (\bar{\omega},\, \bar{k})=(\bar{\omega}_\star,\, \bar{k}_\star) + \epsilon (\delta \bar{\omega},\, \delta \bar{k})\,, \qquad
    s=\frac{\delta \bar{\omega}}{\delta \bar{k}}\,,
\end{equation}
where $\epsilon$ is infinitesimal. Expanding the matrix $\mathcal{M}$ as
\begin{align} \small
\label{eq:MatrixEqnSingAnal}
\begin{split}
\mathcal{M} \simeq & \sum_{l} \mathcal{M}^{(l)} (\epsilon \delta\bar{k})^l\,,
\end{split}
\end{align}
we find the additional conditions, such as $\mathcal{M}^{(l)}\Psi=0$ for $l<0$. Examining all the relations between coefficient $\{Z^{(i)},\, E^{(i)},\, \Phi^{(i)} \}$ with an additional free parameter $s$, the point $(\bar{\omega}_\star, \bar{k}_\star)$ can be  determined whether it is a pole-skipping point or not.

\subsection{Results} \label{sec:Results}
\subsubsection{Spin-$0$ channel}
\begin{figure}
     \centering
     \subfigure[]{\includegraphics[width=0.45\textwidth]{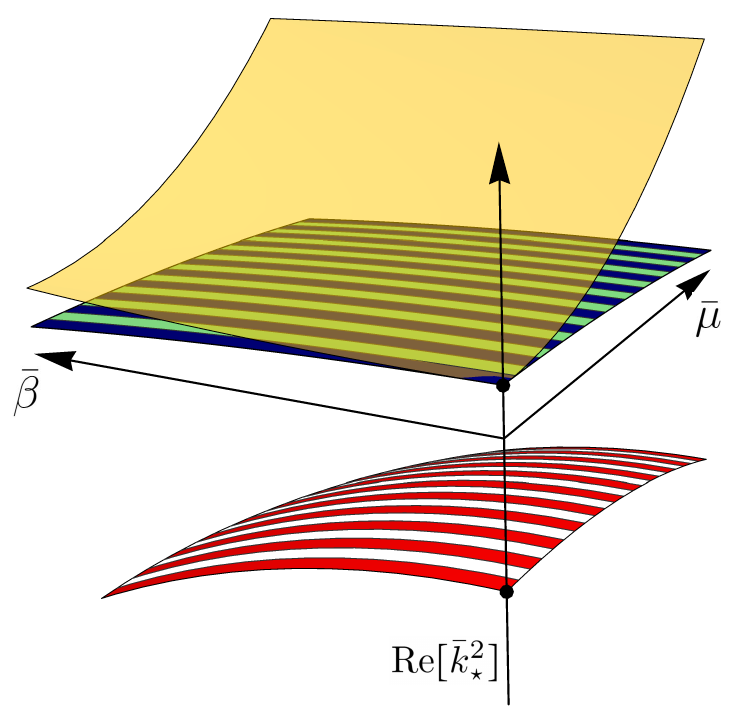}}
     \subfigure[]{\includegraphics[width=0.45\textwidth]{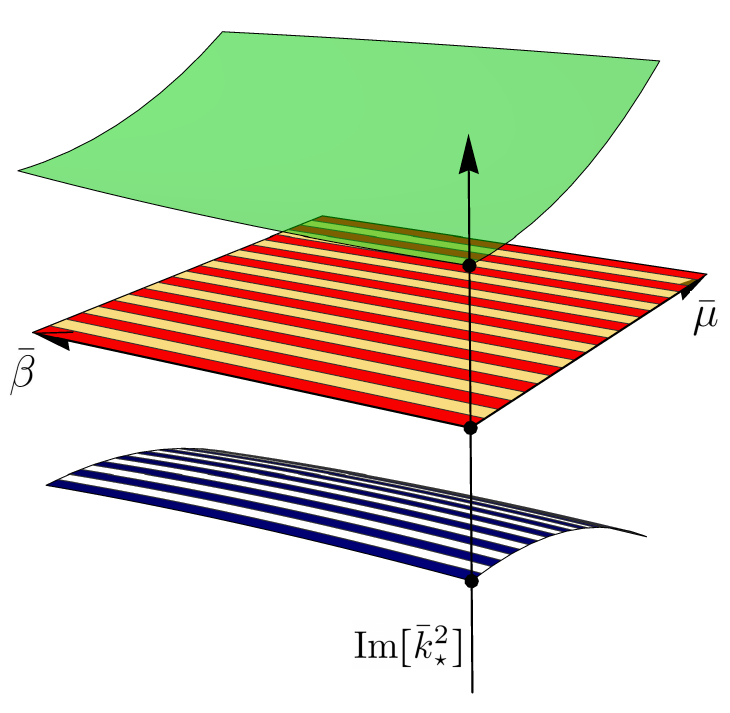}}
     \caption{ (a) Re$[\bar{k}_\star^2]$ and (b) Im$[\bar{k}_\star^2]$, corresponding to solutions of the Eq.~\eqref{eq:MeqnSpin0case1}, as a function of $\bar{\beta}$ and $\bar{\mu}$. The different color patterns (Stripped Red, Yellow, Green, Stripped Blue) correspond to the solutions in Eq.~\eqref{eq:spin0ps}. The black dots correspond to the pole-skipping points in the absence of momentum relaxation ($\bar{\beta}$) and chemical potential ($\bar{\mu}$), $\bar{\beta}=\bar{\mu}=0$.  Both the pure real (Yellow; $\text{Re}[\bar{k}_\star^2]>0$, $\text{Im}[\bar{k}_\star^2]=0$) and pure imaginary (Stripped Red; $\text{Re}[\bar{k}_\star^2]<0$, $\text{Im}[\bar{k}_\star^2]=0$) pole-skipping momentums at $\bar{\beta}=\bar{\mu}=0$ remain to be real, $\text{Im}[\bar{k}_\star^2]=0$, at finite $\bar{\beta}$ and $\bar{\mu}$.}
     \label{fig:spin0}
\end{figure}

\paragraph{Regular cases.}
Using the ansatz \eqref{eq:ZEPhiAnsatz}, the equations of motion \eqref{eq:Spin0EOM} take the form \eqref{eq:NearHorizon Ansatz}, and the matrix equation $\mathcal{M}\Psi=0$ corresponding to near-horizon solutions of \eqref{eq:Spin0EOM} can be obtained. The full tower of pole-skipping points can then be computed by using the condition \eqref{eq:omegaDetM}. For simplicity, we consider \eqref{eq:omegaDetM} for the $n=1$, in which case the pole-skipping frequency is given by $i\bar{\omega}_\star=1$, and the determinant in \eqref{eq:omegaDetM} is given by
\begin{align}\label{eq:MeqnSpin0case1}
\begin{split}
    \left.\det (\mathcal{M}_1)\right|_{(\bar{\omega}_{\star}, \bar{k}_{\star})} \approx &~ \left(4 \bar{k}_\star^2+24 \bar{r}_h^2+\bar{\beta}^2-4 \bar{\mu}^2\right) \bigg\{ 576 \bar{k}_\star^6-48 \left(72 \bar{r}_h^2 -21 \bar{\beta}^2+4 \bar{\mu}^2 \right) \bar{k}_\star^4 \\
    & +4 \left( 6336 \bar{r}_h^4 -48 \bar{r}_h^2 \left(57 \bar{\beta}^2-52 \bar{\mu}^2\right) +243 \bar{\beta}^4 -592 \bar{\mu}^4 -120 \bar{\beta}^2 \bar{\mu}^2 \right) \bar{k}_\star^2 \\
    & -3 \left( 24 \bar{r}_h^2 -3 \bar{\beta}^2-4 \bar{\mu}^2 \right)^2 \left( 24 \bar{r}_h^2 -7 \bar{\beta}^2+44 \bar{\mu}^2 \right) \bigg\}\,.
\end{split}
\end{align}
The equation $\left.\det (\mathcal{M}_1)\right|_{(\bar{\omega}_{\star}, \bar{k}_{\star})}=0$ has eight solutions because $\left.\det (\mathcal{M}_1)\right|_{(\bar{\omega}_{\star}, \bar{k}_{\star})}$ is a polynomial of 8th order in $\bar{k}_{\star}$. This implies that there are eight pole-skipping points at $i\bar{\omega}_\star=1$. Since the exact solutions for any value of $\bar{\mu}$ and $\bar{\beta}$ are exceedingly long and intricate, we compute the pole-skipping points for small values of $\bar{\beta}$ with $\bar{\mu}=0$, or for small values of $\bar{\mu}$ with $\bar{\beta}=0$: 
\begin{align} 
\label{eq:spin0ps}
\begin{split}
\text{Effect of $\bar{\beta}$:}\\
    \left. \bar{k}_\star^2\, \right|_{\bar{\mu}=0} &=
    \begin{cases}
        ~-6 \bar{r}_{h,0}^2 - \frac{7}{4}  \bar{\beta}^2 +\mathcal{O}(\bar{\beta}^4) \,, & \text{(Striped Red)} \\
        ~2 \bar{r}_{h,0}^2 + \frac{1}{4}  \bar{\beta}^2 +\mathcal{O}(\bar{\beta}^4)\,,  & \text{(Yellow)}\\
        ~2 \left(1+2 i \sqrt{2}\right) \bar{r}_{h,0}^2 + \frac{1}{4}\left(-1+i \sqrt{2}\right) \bar{\beta}^2 +\mathcal{O}(\bar{\beta}^4)\,, & \text{(Green)}\\
        ~2 \left(1-2 i \sqrt{2}\right) \bar{r}_{h,0}^2 + \frac{1}{4}\left(-1-i \sqrt{2}\right) \bar{\beta}^2 +\mathcal{O}(\bar{\beta}^4)\,, & \text{(Striped Blue)}
    \end{cases}\\
\text{Effect of $\bar{\mu}$:}\\
    \left. \bar{k}_\star^2\, \right|_{\bar{\beta}=0} &=
    \begin{cases}
        ~ -6 \bar{r}_{h,0}^2 -  \bar{\mu}^2 +\mathcal{O}(\bar{\mu}^4)\,, & \text{(Striped Red)}\\
        ~ 2 \bar{r}_{h,0}^2 + 3 \bar{\mu}^2 +\mathcal{O}(\bar{\mu}^4)\,, & \text{(Yellow)}\\
        ~ 2 \left(1+2 i \sqrt{2}\right) \bar{r}_{h,0}^2 + \frac{1}{3} \left(-1+7i \sqrt{2}\right) \bar{\mu}^2+\mathcal{O}(\bar{\mu}^4)\,, & \text{(Green)}\\
        ~ 2 \left(1-2 i \sqrt{2}\right) \bar{r}_{h,0}^2 + \frac{1}{3} \left(-1-7i \sqrt{2}\right) \bar{\mu}^2+\mathcal{O}(\bar{\mu}^4)\,, & \text{(Striped Blue)}
    \end{cases}
\end{split}
\end{align}
where $\bar{r}_{h,0}={1}/{2}$ denotes the black hole radius given in \eqref{eq:horizonRadius} with $\bar{\beta}=\bar{\mu}=0$.
The real and imaginary parts of the pole-skipping momentum squared $\bar{k}_\star^2$ are also shown in Fig.~\ref{fig:spin0}. Additionally, taking $\bar{\beta}=\bar{\mu}=0$ in \eqref{eq:spin0ps}, we recover the results of ~\cite{Blake:2019otz, Natsuume:2019xcy},\footnote{The results for the sound mode in \eqref{eq:spin0ps1woBR} correspond to a higher-dimensional generalization of the corresponding results obtained in \cite{Blake:2019otz}.} namely:
\begin{align} \label{eq:spin0ps1woBR}
\begin{split}
i\bar{\omega}_\star=1~:\\
    & \bar{k}_\star^2 =
    \begin{cases}
        ~ -6 \bar{r}_{h,0}^2
           & \text{from the axion field}\,, \\
        ~ 2 \bar{r}_{h,0}^2 
           & \text{from the gauge field (scalar mode)}\,, \\
        ~ 2\left( 1 \pm 2 i \sqrt{2} \right) \bar{r}_{h,0}^2 \qquad
           & \text{from the metric field (sound mode)}\,.
    \end{cases}
\end{split}
\end{align}
 
Furthermore, we observed that if the condition $\text{Im}[\bar{k}_\star^2]=0$ is satisfied for $\bar{\beta}=\bar{\mu}=0$, then it remains satisfied for any values of $\bar{\beta}$ and $\bar{\mu}$. In other words, if the pole-skipping momentum $\bar{k}_\star$ is either purely real or purely imaginary for $\bar{\beta}=\bar{\mu}=0$, then it remains purely real or purely imaginary for any values of $\bar{\beta}$ and $\bar{\mu}$: e.g., see the Yellow (and Striped Red) plane in the right panel of Fig. \ref{fig:spin0}. All the other pole-skipping points computed in this manuscript also have this property.

So far, we have computed only the leading pole-skipping points, obtained by solving \eqref{eq:omegaDetM} for $n=1$. A substantial set of subleading pole-skipping points can be computed by solving \eqref{eq:omegaDetM} for larger values of $n$. Nevertheless, we find that this approach proves inadequate in apprehending a specific class of pole-skipping points, which we term as {\it singular pole-skipping points} of gauge-invariant variables. This limitation arises because the components $M_{ij}$ in \eqref{eq:MatrixEqn} take the form\footnote{For all matrix components $M_{ij}$, the denominator takes three possible forms: 
\begin{equation*}
    \bar{\omega} \,,\,\, \bar{k}^2\,,\,\,\,\, \text{or} \,\,\,\, \bar{\omega} \big(36 \bar{r}_h^2 \bar{\omega}^2 - \bar{k}^2 \big( 24 \bar{r}_h^2 -3 \bar{\beta}^2 - 4\bar{\mu}^2 \big)\big).
\end{equation*}
}
\begin{align} \label{eq:MijExample}
\begin{split}
    M_{11} \propto \frac{1}{\bar{\omega} \left( 36 \bar{r}_h^2 \bar{\omega}^2 - \bar{k^2}\left( 24 \bar{r}_h^2 -3 \bar{\beta}^2 - 4\bar{\mu}^2 \right) \right)}\,, \qquad M_{12}\propto\frac{1}{\bar{\omega}}\,, \qquad M_{13}\propto\frac{1}{\bar{k}^2}\,,\\
    M_{21}\propto\frac{1}{\bar{\omega} \left( 36 \bar{r}_h^2 \bar{\omega}^2 - \bar{k}^2 \left( 24\bar{r}_h^2 - 3 \bar{\beta}^2 - 4 \bar{\mu}^2 \right) \right)}\,, \qquad M_{22}\propto\frac{1}{\bar{\omega}}\,, \qquad M_{23}\propto \frac{1}{\bar{k}^2}\,,
\end{split}
\end{align}
which diverges for some specific values of $\bar{\omega}$ and $\bar{k}$.
When $M_{ij}$ diverges as $1/\epsilon$, we first expand the matrix $\mathcal{M}$ in powers of $\epsilon$, namely, $\mathcal{M} \simeq \sum_{l} \mathcal{M}^{(l)} \epsilon^l$, then we find additional pole-skipping points using the relation $\mathcal{M}^{(l)}\Psi=0$ for $l\leq0$, instead of the relation $\mathcal{M}\Psi=0$ in \eqref{eq:MatrixEqn}. Note that it is imperative to investigate this singular case not only as because its omission results in the exclusion of a substantial subset of pole-skipping points, but also to obtain a better understanding of the implications of the pole-skipping phenomenon.

\paragraph{Singular case.}
The components $M_{11}$ and $M_{21}$ in \eqref{eq:MijExample} diverge under the condition:\footnote{The other singular cases of the spin-$0$ channel are analyzed in Appendix \ref{sec:zero0}. These cases do not give any additional pole-skipping points, for finite chemical potential and momentum relaxation.}
\begin{equation} \label{eq:Spin0case2Condition}
    36 \bar{r}_h^2 \bar{\omega}^2 - \bar{k}^2 \left( 24 \bar{r}_h^2 -3 \bar{\beta}^2 - 4\bar{\mu}^2 \right) =0\,, \qquad \bar{\omega} \neq 0\,, \qquad \bar{k} \neq 0\,.
\end{equation}
Using \eqref{eq:singp-sAnalDef}, we expand the matrix $\mathcal{M}$ in \eqref{eq:MatrixEqnSingAnal} with a function $\bar{k}^2_\star(\bar{\omega}^2_\star)$ satisfying the condition \eqref{eq:Spin0case2Condition}:
\begin{align} \small
\label{eq:MatrixEqnSpin0Sing1}
\begin{split}
\mathcal{M} \simeq &
\begin{pmatrix}
M_{11}^{(-1)}& & \\
M_{21}^{(-1)}& & \cdots \\
& \\
\vdots & & \ddots  
\end{pmatrix}
\left(i \bar{\omega}_\star+1 \right)(\epsilon \delta \bar{k})^{-1}
+
\begin{pmatrix}
M_{11}^{(0)} & M_{12}^{(0)} & M_{13}^{(0)} & i\bar{\omega}_\star-1 & & & \\
M_{21}^{(0)} & M_{22}^{(0)} & M_{23}^{(0)} & & i \bar{\omega}_\star-1 & & \cdots \\
& & M_{33}^{(0)} & & & i \bar{\omega}_\star-1 & \\
    &   & \vdots  & & &  & \ddots
\end{pmatrix}
+\mathcal{O}(\epsilon \delta \bar{k})\,,    
\end{split}
\end{align}
where the coefficients $M_{ij}^{(l)}$ depend on $\bar{\omega}_\star$ and $s$. This allows us to find additional conditions, such as $\mathcal{M}^{(-1)}\Psi=0$. By using these relations, all the coefficients $\{Z_z^{(i)},\, E_z^{(i)},\, \Phi_z^{(i)} \}$ in \eqref{eq:ZEPhiAnsatz} are determined in terms of seven parameters $\{ Z_z^{(0)},\, E_z^{(0)},\, \Phi_z^{(0)},\, Z_z^{(1)},\, E_z^{(1)},\, \Phi_z^{(1)}, s \}$. However, \eqref{eq:MatrixEqnSpin0Sing1} also gives four constraints involving the same parameters, which means three of these coefficients can be taken as free parameters. This implies that the gauge invariant variables $Z_z$, $E_z$ and $\Phi_z$ are uniquely determined, up to an overall factor.
However, at $i\bar{\omega}_\star=-1$, one constraint is lost, and all the coefficients are determined by four free parameters, that can be taken as\footnote{The coefficient $Z_z^{(1)}$ can be written in terms of $Z_z^{(0)}$, $E_z^{(0)}$, and $\Phi_z^{(0)}$:
\begin{align} \label{eq:Spin0SingZ1Para} \small
    Z_z^{(1)} =
    \frac{3}{4 \pi T \bar{r}_h}
    \frac{\left(6 \bar{r}_h^{3/2}-7 \bar{r}_h^{\frac{1}{2}}\right)s -4 \sqrt{3} \bar{r}_h+3 \sqrt{3}}{3 \bar{r}_h^{\frac{1}{2}} s - \sqrt{3}} Z_z^{(0)}-\frac{2 \bar{\mu}}{\bar{r}_h} E_z^{(0)} +\frac{i \bar{\beta}}{24 \pi^2 T^2 \bar{r}_h^3} \Phi_z^{(0)}\,,
\end{align}
but it also depends on the slope $s$, which is a free parameter. That implies that $Z_z^{(1)}$ can be taken as a free parameter.} $Z_z^{(0)}$, $E_z^{(0)}$, $\Phi_z^{(0)}$, and $Z_z^{(1)}$. This implies that the solution is no longer unique, and the point in question is a pole-skipping point of the spin-0 sector.

Finally, using \eqref{eq:Spin0case2Condition}, the pole-skipping momentum at $i\bar{\omega}_\star=-1$ can be written as
\begin{equation} \label{eq:leadingPS}
   \bar{k}_\star^2 = -6 \bar{r}_h^2+\frac{3}{4} \bar{\beta}^2 + \bar{\mu}^2\,,
\end{equation}
where taking $\bar{\beta}=\bar{\mu}=0$, it recovers the result of~\cite{Blake:2018leo}:
\begin{align}\label{lppsm}
\begin{split}
i\bar{\omega}_\star=-1~:\\
    & \bar{k}_\star^2 =
    -6 \bar{r}_{h,0}^2 \qquad \qquad \text{metric field (sound mode)}\,.
\end{split}
\end{align}

It has been proposed that the leading pole-skipping point in the spin-0 channel is related to the Lyapunov exponent and butterfly velocity as follows~\cite{Grozdanov:2017ajz, Blake:2017ris}
\begin{equation}\label{eq:PSchaos}
    \omega_\star=i \lambda_L\,, \qquad k_\star^2=-\frac{\lambda_L^2}{v_B^2}\,.
\end{equation}
where the Lyapunov exponent and butterfly velocity associated with the action \eqref{eq:action} are given by~\cite{Abbasi:2020ykq}:
\begin{equation} \label{eq:lv}
    \lambda_L=2\pi T\,, \qquad v_B^2 = \frac{\lambda_L^2 r_h}{{6 \pi T}} \,.
\end{equation}
Using \eqref{eq:Temp}, we checked that this relation is still satisfied in the presence of $\mu$ and $\beta$.

\subsubsection{Spin-$1$ channel} \label{sec:spin1}
\begin{figure}
    \centering
    \includegraphics[width=0.45\textwidth]{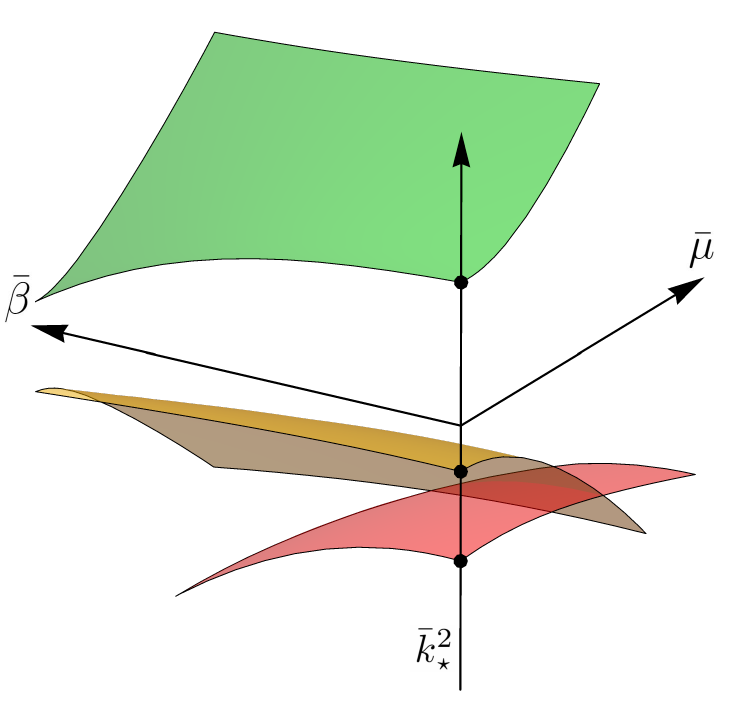}
    \caption{Solutions of the Eq.~\eqref{eq:spin1ps1} in terms of $\bar{k}_\star^2$. The different colors (Red, Yellow, Green) correspond to the solution in Eq.~\eqref{eq:spin1ps}. The black points represent the pole-skipping points in the absence of momentum relaxation ($\bar{\beta}$) and chemical potential ($\bar{\mu}$), $\bar{\beta}=\bar{\mu}=0$. In this figure, all the solutions $\bar{k}_\star^2$ are real for finite $\bar{\beta}$ and $\bar{\mu}$, implying that both the pure real (Green; $\bar{k}_\star^2>0$) and pure imaginary (Red, Yellow; $\bar{k}_\star^2<0$) pole-skipping momentums $\bar{k}_\star^2$ at $\bar{\beta}=\bar{\mu}=0$ remain to be real, $\text{Im}[\bar{k}_\star^2]=0$, at finite $\bar{\beta}$ and $\bar{\mu}$.}
    \label{fig:spin1}
\end{figure}

\paragraph{Regular case.}
The analysis for the spin-1 channel is similar to the analysis for the spin-0 channel. We first expand the equations \eqref{eq:Spin1EOM} into the near-horizon region, obtaining equations of the form \eqref{eq:NearHorizon Ansatz}. Then we find the matrix equation $\mathcal{M}\Psi=0$ as in \eqref{eq:MatrixEqn}. The pole-skipping points for $n=1$ can then be obtained from the determinant in \eqref{eq:omegaDetM} which is
\begin{align}
\begin{split}
     \left.\det (\mathcal{M}_1)\right|_{(\bar{\omega}_{\star}, \bar{k}_{\star})} \approx &~
     \left(4 \bar{k}_\star^2+24 \bar{r}_h^2+\bar{\beta}^2-4 \bar{\mu}^2\right)
     \bigg\{48 \bar{k}_\star^4 -8 \left(24 \bar{r}_h^2-9 \bar{\beta}^2-4 \bar{\mu}^2\right) \bar{k}_\star^2\\
     & -\left(24 \bar{r}_h^2 -3\bar{\beta}^2 -4\bar{\mu}^2\right) \left(24 \bar{r}_h^2 -7\bar{\beta}^2 +44\bar{\mu}^2\right) \bigg\}\,.
\end{split}
\end{align}
Therefore, the pole-skipping points for $i\bar{\omega}_\star=1$ take the form
\begin{align} \label{eq:spin1ps1}
\begin{split}
    \bar{k}_\star^2 = 
 \begin{cases}
    ~ -6\bar{r}_h^2-\frac{1}{4}\bar{\beta}^2+\bar{\mu}^2 \,,\\
    ~  2\bar{r}_h^2-\frac{3}{4}\bar{\beta}^2-\frac{\bar{\mu}^2}{3} \pm \frac{1}{3}\sqrt{9 \left( 4 \bar{r}_h^2 - \bar{\beta}^2 \right)^2 +3 \left(56 \bar{r}_h^2 -5 \bar{\beta}^2 \right)\bar{\mu}^2 - 32 \bar{\mu}^4}\,.
 \end{cases}
\end{split}
\end{align}
For small values of $\bar{\beta}$ and $\bar{\mu}$, these pole-skipping points take the form
\begin{align} 
\label{eq:spin1ps}
\begin{split}
\text{Effect of $\bar{\beta}$:}\\
    & \left. \bar{k}_\star^2\, \right|_{\bar{\mu}=0} =
    \begin{cases}
        ~ -6\bar{r}_h^2-\frac{1}{4}\bar{\beta}^2\,, \qquad & \text{(Red)} \\
        ~ -2 \bar{r}_h^2 + \frac{1}{4} \bar{\beta}^2 +\mathcal{O}(\bar{\beta}^4)\,, \qquad & \text{(Yellow)}\\
        ~ 6 \bar{r}_h^2 -\frac{7}{4} \bar{\beta}^2 +\mathcal{O}(\bar{\beta}^4)\,,  \qquad & \text{(Green)}\\
    \end{cases}\\
\text{Effect of $\bar{\mu}$:}\\
    & \left. \bar{k}_\star^2\, \right|_{\bar{\beta}=0} =
    \begin{cases}
        ~ -6\bar{r}_h^2 +\bar{\mu}^2\,, \qquad & \text{(Red)}\\
        ~ -2 \bar{r}_h^2 -\frac{8}{3} \bar{\mu}^2 +\mathcal{O}(\bar{\mu}^4)\,,  \qquad & \text{(Yellow)}\\
        ~ 6 \bar{r}_h^2 + 2 \bar{\mu}^2 +\mathcal{O}(\bar{\mu}^4)\,. \qquad & \text{(Green)}
    \end{cases}
\end{split}
\end{align}
The behavior of $\bar{k}^2$ as a function of $\bar{\beta}$ and $\bar{\mu}$ is shown in Fig. \ref{fig:spin1}.
At $\bar{\beta}=\bar{\mu}=0$, our results are consistent with the previous literature in~\cite{Blake:2019otz, Natsuume:2019xcy},\footnote{The pole-skipping points for the metric field (shear mode) in \eqref{eq:spin1ps1woBR} are consistent with the one from a higher-dimensional generalization of \cite{Blake:2019otz}.} namely:
\begin{align}\label{eq:spin1ps1woBR}
\begin{split}
i\bar{\omega}_\star=1~:\\
    & \bar{k}_\star^2 =
    \begin{cases}
        ~ -6 \bar{r}_{h,0}^2 \qquad
           & \text{from the axion field}\,, \\
        ~ -2 \bar{r}_{h,0}^2 \qquad
           & \text{from the gauge field (vector mode)}\,, \\
        ~ 6 \bar{r}_{h,0}^2 \qquad
           & \text{from the metric field (shear mode)}\,.
    \end{cases}
\end{split}
\end{align}
Using the same approach explained above, the remaining pole-skipping points corresponding to higher values $n$ can be computed. However, as in the previous case, it is important to note that there exists pole-skipping points that cannot be found by this method due to singularity issues. This is primarily due to the denominators in the components $M_{ij}$ of \eqref{eq:MatrixEqn}, such as
\begin{align} \label{eq:Spin1Denomi}
    M_{11} \propto \frac{1}{\bar{\omega}}\,, \qquad M_{13} \propto \frac{1}{\bar{k}}\,, \qquad M_{21} \propto \frac{1}{\bar{\omega}}\,, \qquad M_{23} \propto \frac{1}{\bar{k}}\,,
\end{align}
which diverges for $\bar{\omega}=0$ and/or for $\bar{k}=0$. Now we proceed to consider these singular cases following the same approach we used in the last section.

\paragraph{Singular case.}
The components $M_{11}$, and $M_{21}$ in \eqref{eq:Spin1Denomi} diverge when:\footnote{We also examine other singular cases for the spin-$1$ channel. These are analyzed in Appendix \ref{sec:zero1}. However, these singular cases do not give rise to any additional pole-skipping points.}
\begin{equation} \label{eq:Spin1Sing1Condition}
    \bar{\omega}=0\,, \qquad \bar{k}\neq0\,.
\end{equation}
Let us denote the point satisfying \eqref{eq:Spin1Sing1Condition} as $(\bar{\omega}_\star,\bar{k}_\star)$. Using \eqref{eq:singp-sAnalDef} to compute $\mathcal{M}$ in a point infinitesimally close to $(\bar{\omega}_\star,\bar{k}_\star)$, we obtain:
\begin{align} \small
\label{eq:MatrixEqnSpin1Sing1}
\begin{split}
\mathcal{M} \simeq
\begin{pmatrix}
M_{11}^{(-1)} & & \\
M_{21}^{(-1)} & & \cdots \\
& & \\
\vdots & & \ddots  
\end{pmatrix}
\left( \bar{k}_\star^2+\bar{\beta}^2 \right) (\epsilon \delta \bar{k})^{-1}
+
\begin{pmatrix}
M_{11}^{(0)} & M_{12}^{(0)} & M_{13}^{(0)} & -1 & & & \\
M_{21}^{(0)} & M_{22}^{(0)} & M_{23}^{(0)} & & -1 & & \cdots \\
& & M_{33}^{(0)} & & & -1 & \\
    &   & \vdots  & & &  & \ddots
\end{pmatrix}
+\mathcal{O}(\epsilon \delta \bar{k})\,,
\end{split}
\end{align}
where $M_{ij}$ depends on $\bar{k}_\star$ and $s$. In general, all coefficients $\{Z_\alpha^{(i)},\, E_\alpha^{(i)},\, \Phi_\alpha^{(i)} \}$ can be determined by seven parameters, $\{ Z_\alpha^{(0)},\, E_\alpha^{(0)},\, \Phi_\alpha^{(0)},\, Z_\alpha^{(1)},\, E_\alpha^{(1)},\, \Phi_\alpha^{(1)}, s \}$, which are subjected to four constraints. Therefore, all the coefficients can be determined by three free parameters, and the gauge invariant quantities $Z_{\alpha}$, $E_{\alpha}$ and $\Phi_{\alpha}$  are uniquely determined up to an overall constant.
However, at $\bar{k}_\star=\pm i\bar{\beta}$, we loose one constrain, and the coefficients are determined by four free parameters, which can be taken as \{$Z_\alpha^{(0)}$, $E_\alpha^{(0)}$, $\Phi_\alpha^{(0)}$, and $Z_\alpha^{(1)}$\}, implying that the solutions are no longer unique, so that we find the pole-skipping as
\begin{align}\label{NEPSP}
\begin{split}
(\bar{\omega}_\star,\,\bar{k}_\star)=(0,\,\pm i\bar{\beta})  \,.
\end{split}
\end{align}
%

One intriguing remark for \eqref{NEPSP} is in order. 
The leading pole-skipping point \eqref{NEPSP} seems to be a new pole-skipping phenomenon emerging at finite $\bar{\beta}$, distinct from the $(\bar{\omega}_\star,~ \bar{k}_\star) = \left(0, 0 \right)$ pole-skipping point
reported in~\cite{Blake:2019otz, Natsuume:2019xcy,Kovtun:2012rj}.\footnote{Refer to Appendix \ref{sec:zero0} for a further discussion on the $\left(0, 0 \right)$ case.} Its nature and origin can be grasped from one of the equations within \eqref{eq:MatrixEqnSpin1Sing1}:
\begin{align}\label{NEWEOM}
\begin{split}
Z_{\alpha}^{(1)} - \left( \frac{i \bar{k}_\star}{\pi \bar{r}_h^2 s T} + \frac{\bar{k}_\star^2+\bar{\beta}^2}{4 \pi \bar{r}_h^2} \right) Z_{\alpha}^{(0)}
+ \frac{2\bar{\mu}}{\bar{r}_h} E_{\alpha}^{(0)}
+ \frac{\bar{\beta}}{4\pi^2 \bar{r}_h^2 T^2 \bar{k}_{\star}} \Phi_{\alpha}^{(0)}= 0 \,.
\end{split}
\end{align}
Upon setting $\bar{\beta}=0$, it is evident that $\Phi_{\alpha}^{(0)}$ uncouples from other fields as it should.\footnote{Recall that when $\beta=0$, the axion fluctuation ($\Phi$) function as the probe field and thus remain independent of other fields.} However, the imposition of $\bar{k}_{\star} = \pm i \bar{\beta}$, the pole-skipping condition, alters this scenario, as illustrated by the persistence of coupling in the last term of \eqref{NEWEOM}: $\Phi_{\alpha}^{(0)}$ remains coupled with other fields even at $\bar{\beta}=0$. Consequently, $\bar{k}_{\star} = \pm i \bar{\beta}$ stands as a reliable condition solely for finite $\bar{\beta}$ cases.\footnote{At $\bar{k}_{\star} = \pm i \bar{\beta}$, \eqref{NEWEOM} simplifies to:
\begin{align}\label{eq:Spin1SingZ1Para} 
Z_\alpha^{(1)} = \mp \frac{\bar{\beta}}{\pi  \bar{r}_h^2 s T} Z_\alpha^{(0)} -\frac{2 \bar{\mu}}{\bar{r}_h} E_\alpha^{(0)} \pm \frac{i}{4 \pi ^2 \bar{r}_h^2 T^2} \Phi_\alpha^{(0)} \,,
\end{align}
where one can notice that \eqref{NEPSP} corresponds to a pole-skipping point, as the free parameter $s$ appears in \eqref{eq:Spin1SingZ1Para}.} As such, our newly identified pole-skipping point \eqref{NEPSP} manifests only at finite $\bar{\beta}$.\footnote{We may change the way to approach the pole-skipping points in \eqref{eq:singp-sAnalDef}, which may affect this conclusion. We leave this issue as a future work.}

\subsubsection{Spin-2 channel}
In this case, the equations of motion are decoupled~\eqref{eq:Spin2EOM}, and we can obtain them in the matrix form following the procedure outline in \cite{Blake:2019otz}. We obtain
\begin{equation} \label{eq:MatrixEqnSpin2}
    \tilde{\mathcal{M}} Z_{\alpha\beta} \equiv
    \begin{pmatrix}
    \tilde{M}_{11} & i \bar{\omega}-1 & 0 & 0 &  \\
    \tilde{M}_{21} & \tilde{M}_{22} & i \bar{\omega}-2 & 0 &\cdots \\
    \tilde{M}_{31} & \tilde{M}_{32} & \tilde{M}_{33} & i \bar{\omega}-3 & \\
    & \vdots & & & \ddots
    \end{pmatrix}
    \begin{pmatrix}
    Z_{\alpha\beta}^{(0)} \\
    Z_{\alpha\beta}^{(1)} \\
    Z_{\alpha\beta}^{(2)} \\
    Z_{\alpha\beta}^{(3)} \\
    \vdots
    \end{pmatrix}
    =
    0\,.
\end{equation}
The pole-skipping points $(\bar{\omega}_\star,\bar{k}_\star)$ satisfy the condition:
\begin{equation} \label{eq:omegaDetMSpin2}
    i \bar{\omega}_{\star}= n, \qquad \left.\det (\tilde{\mathcal{M}}_n)\right|_{(\bar{\omega}_{\star}, \bar{k}_{\star})}=0\,,
\end{equation}
where $n$ is a natural number, and $\tilde{\mathcal{M}}_n$ represents a square matrix obtained from the matrix $\tilde{\mathcal{M}}$ in \eqref{eq:MatrixEqnSpin2} up to the $n$-th column. For $n=1$, the determinant in \eqref{eq:omegaDetMSpin2} reads
\begin{equation}
    \left.\det(
    \tilde{\mathcal{M}}_1)
    \right|_{(\bar{\omega}_{\star}, \bar{k}_{\star})} \approx 4 \bar{k}_\star^2 + 24 \bar{r}_h^2 + \bar{\beta}^2 - 4 \bar{\mu}^2 \,.
\end{equation}
Therefore, the pole-skipping points for $i\bar{\omega}_\star=1$ are given by
\begin{equation} \label{eq:Spin2ps1}
    \bar{k}_\star^2 = -6 \bar{r}_h^2 - \frac{\bar{\beta}^2}{4} + \bar{\mu}^2\,,
\end{equation}
which, at $\bar{\beta}=\bar{\mu}=0$, reproduce the result in~\cite{Natsuume:2019xcy}:
\begin{align} \label{eq:Spin2ps1woBR}
\begin{split}
i\bar{\omega}_\star=1~:\\
    & \bar{k}_\star^2 = -6 \bar{r}_{h,0}^2 \qquad \qquad \text{from the metric field (tensor mode)}\,.
\end{split}
\end{align}

The matrix elements of $\tilde{\mathcal{M}}$ do not diverge for any values of $\bar{\omega}$ and $\bar{k}$. Therefore, unlike the spin-0 and spin-1 sectors, there are no singular pole-skipping points for the spin-2 sector.

\section{Conclusion}\label{sec:discussion}
In this work, we studied the pole-skipping phenomenon in holographic axion theories in five dimensions, which is the typical class of holographic framework for studying strongly coupled systems at finite chemical potential ($\mu$) and momentum relaxation ($\beta$). 

It is worth noting that prior studies primarily employed the near-horizon method to identify pole-skipping points within relatively simple backgrounds (e.g., $\mu=0$ and/or $\beta=0$), featuring equations of motion of moderate complexity. This study expands upon the near-horizon method to address coupled equations of motion when $\mu\neq0$ and/or $\beta\neq0$.

At the linear response level, bulk fluctuations in the metric, gauge, and axion field can be classified into three distinct sectors corresponding to spin-0, spin-1, and spin-2 channels. In addition, the inclusion of backreaction in holographic axion theories, characterized by finite $\mu$ and $\beta$, results in coupled equations of motion. Consequently, all fluctuations of metric, gauge, and axion fields become mutually {\it coupled} within their respective channels.

To systematically address coupled equations of motion, we employ {\it gauge-invariant variables} and investigate pole-skipping points within each sector. Specifically, we utilize the near-horizon method, as introduced in~\cite{Blake:2018leo, Blake:2019otz}, where solutions to the fluctuation equations are sought in the vicinity of the horizon, adhering to ingoing boundary conditions at the horizon. The identification of non-unique solutions serves as a criterion for finding pole-skipping points.

We show that within the framework of gauge-invariant variables, pole-skipping points can be further characterized into two classes: {\it regular} (i.e., conventional) pole-skipping points and {\it singular} pole-skipping points. Regular(singular) pole-skipping point is defined as a pole-skipping point whose equations of motion for gauge-invariant variables are regular(singular or ill-defined). Especially, we observe that all pole-skipping points in the lower-half plane  ($i\bar{\omega}_\star>0$) are identified as regular pole-skipping points.

It is also worth mentioning that within the spin-0 and spin-1 channels, the leading pole-skipping points are singular pole-skipping points. They are not located in the lower-half plane:
\begin{align} \label{eq:summary}
\small
\begin{split} 
& \quad (\bar{\omega}_\star, \bar{k}_\star^2)\big|_{\text{leading}} =
\begin{cases}
    ~ \Big( +i,~ -6 \bar{r}_h^2+\frac{3}{4} \bar{\beta}^2 + \bar{\mu}^2 \Big) = (i,~-1/v_B^2)\qquad \qquad  &\text{(spin-$0$ channel)}\,,\\
    ~ \Big(\,\,\,\,\,\, 0,~ -\bar{\beta}^2 ~ \big) \qquad &\text{(spin-$1$ channel)}\,,\\
    ~ \Big( -i,~ -6 \bar{r}_h^2 - \frac{1}{4}\bar{\beta}^2 + \bar{\mu}^2 \Big) \qquad \qquad  &\text{(spin-$2$ channel)}\,,
    \end{cases}
\end{split}
\end{align}
where $\bar{\omega}_\star=\omega_\star/\lambda_L$ and $\bar{k}_\star=k_\star/\lambda_L$: see also \eqref{eq:lv}.
This implies that the pole-skipping point associated with quantum chaos, as indicated by the Lyapunov exponent ($\lambda_L$) and butterfly velocity ($v_B$), corresponds to a singular pole-skipping points.\footnote{Note that the result in the spin-0 channel \eqref{eq:summary} align with findings obtained without the utilization of gauge-invariant variables in~\cite{Jeong:2021zhz}.} 
In the spin-2 channel, the leading pole-skipping points are regular pole-skipping points. These are located in the lower-half plane.  Compared to the previous research we also found the new sub-leading pole-skipping points.

Two remarks are in order.
First, we for the first time extended the methodology of the near-horizon analysis involving only one equation to coupled equations built from gauge-invariant variables. Furthermore, we developed the near-horizon analysis for the singular pole-skipping points. These generalizations of the methodology will be useful for a more complete analysis and various physical situations. 
Second, as the effect of $\mu$ and $\beta$, we observe that if the pole-skipping momentum $\bar{k}_\star$ is purely real or imaginary for $\bar{\beta}=\bar{\mu}=0$, it maintains this property for any values of $\bar{\beta}$ and $\bar{\mu}$. See for instance Figs. \ref{fig:spin0} and \ref{fig:spin1}. All pole-skipping points identified in this study exhibit this attribute.
Exploration of these phenomena and their physical implications merits further investigation in the future.

\acknowledgments
This work was supported by the Basic Science Research Program through the National Research Foundation of Korea (NRF) funded by the Ministry of Science, ICT $\&$ Future Planning (NRF-2021R1A2C1006791) and GIST Research Institute(GRI) grant funded by the GIST in 2024. This work was also supported by Creation of the Quantum Information Science R$\&$D Ecosystem (Grant No. 2022M3H3A106307411) through the National Research Foundation of Korea (NRF) funded by the Korean government (Ministry of Science and ICT).
Y.A acknowledges the support of the Shanghai Municipal Science and Technology Major Project (Grant No.2019SHZDZX01).
H.-S Jeong acknowledges the support of the Spanish MINECO ``Centro de Excelencia Severo Ochoa'' Programme under grant SEV-2012-0249. This work is supported through the grants CEX2020-001007-S and PID2021-123017NB-I00, funded by MCIN/AEI/10.13039/501100011033 and by ERDF A way of making Europe.
V.~Jahnke and M.~Nishida were supported by the Basic Science Research Program through the National Research Foundation of Korea (NRF) funded by the Ministry of Education (
NRF-2020R1I1A1A01073135, RS-2023-00245035).
All authors contributed equally to this paper and should be considered as co-first authors.

\appendix

\section{Other singular cases}
\label{sec:zero}
This appendix considers all the other possible singular cases from Eq.~\eqref{eq:MijExample} and Eq.~\eqref{eq:Spin1Denomi}. For finite $\mu$ and $\beta$, the following singular cases do not lead to additional pole-skipping points.

\subsection{Spin-$0$ channel} \label{sec:zero0}

\paragraph{Case \uppercase\expandafter{\romannumeral1}.} 
The components $M_{11}$, $M_{12}$, $M_{13}$, $M_{21}$, $M_{22}$, and $M_{23}$ in \eqref{eq:MijExample} diverge when:
\begin{equation} \label{eq:Spin0Sing2Condition}
    \bar{\omega}=0\,, \qquad \bar{k}=0\,.
\end{equation}
Using \eqref{eq:singp-sAnalDef}, we expand the matrix $\mathcal{M}$ in \eqref{eq:MatrixEqnSingAnal} around \eqref{eq:Spin0Sing2Condition}, obtaining:
\begin{equation} \small
\label{eq:MatrixEqnSpin0Sing2}
\mathcal{M} \simeq
\begin{pmatrix}
M_{11}^{(-1)} & & M_{13}^{(-1)} & & \\
\frac{-\bar{\mu}}{2 r_h} M_{11}^{(-1)} & & \frac{-\bar{\mu}}{2 r_h} M_{13}^{(-1)} & & \cdots \\
& \\
& \vdots & & & \ddots  
\end{pmatrix}
\bar{\beta}(\epsilon \delta \bar{k})^{-1}
+
\begin{pmatrix}
M_{11}^{(0)} & M_{12}^{(0)} & & -1 & & & \\
M_{21}^{(0)} & M_{22}^{(0)} & & & -1 & & \cdots \\
& & M_{33}^{(0)} & & & -1 & \\
    &   & \vdots  & & &  & \ddots
\end{pmatrix}
+\mathcal{O}(\epsilon \delta \bar{k})\,,
\end{equation}
where $M_{ij}$ is a function of $s$. This leads to additional conditions, such as $\mathcal{M}^{(-1)}\Psi=0$, similarly to Eq.~ \eqref{eq:MatrixEqnSpin0Sing1}. All the coefficients $\{Z_z^{(i)},\, E_z^{(i)},\, \Phi_z^{(i)} \}$ in \eqref{eq:ZEPhiAnsatz} are determined by three free parameters, implying that the solutions are unique and the point in question is not a pole-skipping point.

When the momentum relaxation $\bar{\beta}$ vanishes, the leading order term $\mathcal{M}^{(-1)}$ in \eqref{eq:MatrixEqnSpin0Sing2} becomes zero. Then, all the series coefficients $\{Z_z^{(i)},\, E_z^{(i)},\, \Phi_z^{(i)} \}$ are determined by four free parameters, that can be taken as $Z_z^{(0)}$, $E_z^{(0)}$, $\Phi_z^{(0)}$, and $Z_z^{(1)}$, implying that the solutions are not unique, and the point $(\bar{\omega}_\star,\,\bar{k}_\star)=(0,\,0)$ is a pole-skipping point. Note that the $Z_z^{(1)}$ is a free parameter because it depends on the slope $s$ defined in \eqref{eq:singp-sAnalDef}:
\begin{equation}
    Z_z^{(1)}=\frac{2 \left(\bar{\mu}^2+3 \bar{r}_h^2\right)}{\pi  \bar{r}_h T \left(\bar{\mu}^2+\bar{r}_h^2 \left(9 s^2-6\right)\right)} Z_z^{(0)} -\frac{4 \bar{\mu}}{\bar{r}_h} E_z^{(0)}\,.
\end{equation}
At $\bar{\beta}=\bar{\mu}=0$, this pole-skipping point recovers the results obtained in~\cite{Blake:2019otz, Natsuume:2019xcy,Kovtun:2012rj}:
\begin{equation}
    (\bar{\omega}_\star,~ \bar{k}_\star) = \left(0,~ 0 \right) \qquad \text{from the metric field (sound mode)}\,.
\end{equation}
Note, however, that this singular case is not associated with the scalar mode of the gauge field fluctuations computed at $\bar{\beta}=\bar{\mu}=0$ in~\cite{Blake:2019otz, Natsuume:2019xcy}:
\begin{equation} \label{eq:psGFSM}
    (\bar{\omega}_\star,~ \bar{k}_\star) = \left(0,~ 0 \right) \qquad \text{from the gauge field (scalar mode)}\,,
\end{equation}
because our method can not capture pole-skipping point \eqref{eq:psGFSM}.

\paragraph{Case \uppercase\expandafter{\romannumeral2}.}
The components $M_{13}$ and $M_{23}$ in \eqref{eq:MijExample} diverge when:
\begin{equation} \label{eq:Spin0Sing4Condition}
    \bar{k}=0\,, \qquad \bar{\omega}\neq0\,.
\end{equation}
Using \eqref{eq:singp-sAnalDef}, we expand the matrix $\mathcal{M}$ in \eqref{eq:MatrixEqnSingAnal} around a point of the form \eqref{eq:Spin0Sing4Condition}, obtaining
\begin{align} \footnotesize
\label{eq:MatrixEqnSpin0Sing4}
\begin{split}
\mathcal{M} \simeq
\begin{pmatrix}
& & M_{13}^{(-2)} & \\
& & M_{23}^{(-2)} & \cdots \\
& & & \\
& & \vdots & \ddots  
\end{pmatrix}
\bar{\beta} (\epsilon \delta \bar{k})^{-2}
+
\begin{pmatrix}
& & M_{13}^{(-1)} & \\
& & M_{23}^{(-1)} & \cdots \\
& & & \\
& & \vdots & \ddots  
\end{pmatrix}
s \bar{\beta} (\epsilon \delta \bar{k})^{-1}
+
\begin{pmatrix}
M_{11}^{(0)} & M_{12}^{(0)} & & i\bar{\omega}_\star-1 & & & \\
M_{21}^{(0)} & M_{22}^{(0)} & & & i\bar{\omega}_\star-1 & & \cdots \\
& & M_{33}^{(0)} & & & i\bar{\omega}_\star-1 & \\
    &   & \vdots  & & &  & \ddots
\end{pmatrix}
+\mathcal{O}(\epsilon \delta \bar{k})\,,
\end{split}
\end{align}
where the matrix components $M_{ij}$ depend on $\bar{\omega}_\star$ and $s$. This gives rise to additional conditions such as $\mathcal{M}^{(-2)}\Psi=0$ and $\mathcal{M}^{(-1)}\Psi=0$. All the coefficients in \eqref{eq:ZEPhiAnsatz} are determined by three free parameters because it is impossible to decrease the number of constraints by setting a specific value to $\bar{\omega}_\star$. Therefore, the corresponding solutions are unique and there are no pole-skipping points of the form given in \eqref{eq:Spin0Sing4Condition}.

\paragraph{Case \uppercase\expandafter{\romannumeral3}.}
The components $M_{11}$, $M_{12}$, $M_{21}$, and $M_{22}$ in \eqref{eq:MijExample} diverge when:
\begin{equation} \label{eq:Spin0Sing3Condition}
    \bar{\omega}=0\,, \qquad \bar{k}\neq0\,.
\end{equation}
Using \eqref{eq:singp-sAnalDef}, we expand the matrix $\mathcal{M}$ in \eqref{eq:MatrixEqnSingAnal} around a point of the form \eqref{eq:Spin0Sing3Condition}, obtaining:
\begin{align} \small
\label{eq:MatrixEqnSpin0Sing3}
\begin{split}
\mathcal{M} \simeq
\begin{pmatrix}
M_{11}^{(-1)} & M_{12}^{(-1)} & & \\
\mathcal{B} M_{11}^{(-1)} & \mathcal{B} M_{12}^{(-1)} & & \cdots \\
& & & \\
& \vdots &  & \ddots  
\end{pmatrix}
(\epsilon \delta \bar{k})^{-1}
+
\begin{pmatrix}
M_{11}^{(0)} & M_{12}^{(0)} & & -1 & & & \\
M_{21}^{(0)} & M_{22}^{(0)} & & & -1 & & \cdots \\
& & M_{33}^{(0)} & & & -1 & \\
    &   & \vdots  & & &  & \ddots
\end{pmatrix}
+\mathcal{O}(\epsilon \delta \bar{k})\,,
\end{split}
\end{align}
where $\mathcal{B}=-\frac{3}{2 \mu}-\frac{\bar{\mu}}{r_h}$, and the matrix components $M_{ij}$ depend on $\bar{k}_\star$ and $s$.
All the coefficients in \eqref{eq:ZEPhiAnsatz} are determined by three free parameters because it is impossible to decrease the number of constraints by setting in a specific value at $\bar{k}_\star$. Therefore, the solutions are unique, and there are no pole-skipping point of the form given in \eqref{eq:Spin0Sing3Condition}.

\subsection{Spin-$1$ channel} \label{sec:zero1}

\paragraph{Case \uppercase\expandafter{\romannumeral1}.}
The components $M_{11}$, $M_{13}$, $M_{21}$, and $M_{23}$ in \eqref{eq:Spin1Denomi} diverge when:
\begin{equation} \label{eq:Spin1Sing3Condition}
    \bar{\omega}=0\,, \qquad \bar{k}=0\,.
\end{equation}
Using \eqref{eq:singp-sAnalDef}, we expand $\mathcal{M}$ in \eqref{eq:MatrixEqnSingAnal} around \eqref{eq:Spin1Sing3Condition}, obtaining:
\begin{align} \small
\label{eq:MatrixEqnSpin1Sing3}
\begin{split}
\mathcal{M} \simeq
\begin{pmatrix}
M_{11}^{(-1)} & & M_{13}^{(-1)} & \\
M_{21}^{(-1)} & & M_{23}^{(-1)} & \cdots \\
& & & \\
& \vdots & & \ddots  
\end{pmatrix}
(\epsilon \delta \bar{k})^{-1}
+
\begin{pmatrix}
M_{11}^{(0)} & M_{12}^{(0)} & & -1 & & & \\
& M_{22}^{(0)} & & & -1 & & \cdots \\
& & M_{33}^{(0)} & & & -1 & \\
    &   & \vdots  & & &  & \ddots
\end{pmatrix}
+\mathcal{O}(\epsilon \delta \bar{k})\,.
\end{split}
\end{align}
where the matrix components $M_{ij}$ depend on $s$. All the coefficients $\{Z_\alpha^{(i)},\, E_\alpha^{(i)},\, \Phi_\alpha^{(i)} \}$ in \eqref{eq:ZEPhiAnsatz} are determined by three free parameters, implying that the solutions are unique, and $(\bar{\omega},\, \bar{k})=(0,\,0)$ is not a pole-skipping point of this sector.

\paragraph{Case \uppercase\expandafter{\romannumeral2}.}
The components $M_{13}$ and $M_{23}$ in \eqref{eq:Spin1Denomi} diverge when:
\begin{equation} \label{eq:Spin1Sing2Condition}
    \bar{k}=0\,, \qquad \bar{\omega}\neq0\,.
\end{equation}
Using \eqref{eq:singp-sAnalDef}, we expand the $\mathcal{M}$ in \eqref{eq:MatrixEqnSingAnal} around a point of the form \eqref{eq:Spin1Sing2Condition}, obtaining:
\begin{align} \small
\label{eq:MatrixEqnSpin1Sing2}
\begin{split}
\mathcal{M} \simeq
\begin{pmatrix}
& & M_{13}^{(-1)} & \\
& & M_{23}^{(-1)} & \cdots \\
& & & \\
& & \vdots & \ddots  
\end{pmatrix}
\bar{\beta}(\epsilon \delta \bar{k})^{-1}
+
\begin{pmatrix}
M_{11}^{(0)} & M_{12}^{(0)} & & i\bar{\omega}_\star-1 & & & \\
M_{21}^{(0)} & M_{22}^{(0)} & & & i\bar{\omega}_\star-1 & & \cdots \\
& & M_{33}^{(0)} & & & i\bar{\omega}_\star-1 & \\
    &   & \vdots  & & &  & \ddots
\end{pmatrix}
+\mathcal{O}(\epsilon \delta \bar{k})\,.
\end{split}
\end{align}
where the matrix components $M_{ij}$ depend on $s$. All the coefficients in \eqref{eq:ZEPhiAnsatz} are determined by three free parameters because it is impossible to decrease the number of constraints by setting in a specific value for $\bar{\omega}_\star$. Therefore, the solutions are unique and the points of the form \eqref{eq:Spin1Sing2Condition} are not pole-skipping points.

\section{Auxiliary functions $H_n$} \label{sec:EOMcoef}
Here we give the explicit expressions of the functions $H_n$ used to write the equations of motion in Sec.~\ref{sec:fluc}. The denominators $H_n$ appearing in \eqref{eq:Spin0EOM} and \eqref{eq:Spin1EOM} are defined as
\begin{flushleft}\leftskip=1cm
$\hspace*{-1cm}
H_1=\,
\omega ^2-k^2 f(r)\,,
$\\
$\hspace*{-1cm}
H_2=\,
-\beta ^2 \omega ^2-\frac{1}{6} k^4 r f'(r)+k^4 (-f(r))+k^2 \omega ^2\,,
$\\
$\hspace*{-1cm}
H_3=\,
\frac{18 r^{11}}{r_h^4} \bigg\{-144 k^6 r^{12} f(r)^3-12 k^2 r^6 f(r)^2 \bigg(k^4 (48 r^6-6 \beta ^2 r^4-8 \mu ^2 r_h^4)+k^2 (12 r^6 (6 \beta ^2-7 \omega ^2)-9 \beta ^4 r^4)-108 \beta ^2 r^6 \omega ^2\bigg)
-f(r) \bigg(k^6 (-24 r^6+3 \beta ^2 r^4+4 \mu ^2 r_h^4)^2-96 k^4 r^6 \omega ^2 (24 r^6-3 \beta ^2 r^4-4 \mu ^2 r_h^4)+36 k^2 r^6 \omega ^2 (r^6 (60 \omega ^2-24 \beta ^2)+3 \beta ^4 r^4+4 \beta ^2 \mu ^2 r_h^4)+1296 \beta ^2 r^{12} \omega ^4\bigg)
+\bigg(k^2 \omega  (24 r^6-3 \beta ^2 r^4-4 \mu ^2 r_h^4)-36 r^6 \omega ^3 \bigg)^2 \bigg\}\,,
$\\
$\hspace*{-1cm}
H_4=\,
r \bigg\{4 k^4 r^2 f(r)+k^4 \bigg(8 r^2-\beta ^2\bigg)-12 k^2 r^2 \omega ^2+12 \beta ^2 r^2 \omega ^2 \bigg\} \bigg\{16 k^4 r^4 f(r)^2
-4 r^2 f(r) \bigg(3 \beta ^2+2 k^2\bigg)\bigg(k^2 (\beta ^2-8 r^2)+12 r^2 \omega ^2\bigg)+\bigg(k^2 (8 r^2-\beta ^2)-12 r^2 \omega ^2\bigg)^2\bigg\}\,,
$\\
$\hspace*{-1cm}
H_5=\,
r_h^4 \bigg\{r \bigg(\omega ^2-k^2 f(r)\bigg) \bigg(4 k^4 r^2 f(r)+k^4 (8 r^2-\beta ^2)-12 k^2 r^2 \omega ^2+12 \beta ^2 r^2 \omega ^2\bigg) \bigg(16 k^4 r^4 f(r)^2-4 r^2 f(r) (3 \beta ^2+2 k^2) (k^2 (\beta ^2-8 r^2)+12 r^2 \omega ^2)+(k^2 (8 r^2-\beta ^2)-12 r^2 \omega ^2)^2\bigg) \bigg(-12 k^4 r^6 f(r)+k^4 (-24 r^6+3 \beta ^2 r^4+4 \mu ^2 r_h^4)+36 k^2 r^6 \omega ^2-36 \beta ^2 r^6 \omega ^2\bigg)\bigg\}\,,
$\\
$\hspace*{-1cm}
H_6=\,
r_h^4 \bigg\{3 k^2 r^7 f(r)+k^2 \bigg(6 r^7-\mu ^2 r r_h^4\bigg)-9 r^7 \omega ^2\bigg\}\,,
$\\
$\hspace*{-1cm}
H_7=\,
\omega ^2-f(r) \bigg\{\beta ^2+k^2\bigg\}\,,
$
\end{flushleft}
\begin{equation}
\end{equation}
where the emblackening factor $f(r)$ is defined as \eqref{eq:givenBackground}.

\bibliographystyle{JHEP}

\providecommand{\href}[2]{#2}\begingroup\raggedright\endgroup

\end{document}